\documentclass[sigconf]{acmart}

\copyrightyear{2026}
\acmYear{2026}
\setcopyright{cc}
\setcctype{by}
\acmConference[KDD 2026] {Proceedings of the 32nd ACM SIGKDD Conference on Knowledge Discovery and Data Mining V.2}{August 9--13, 2026}{Jeju Island, Republic of Korea.}
\acmBooktitle{Proceedings of the 32nd ACM SIGKDD Conference on Knowledge Discovery and Data Mining V.2 (KDD 2026), August 9--13, 2026, Jeju Island, Republic of Korea}
\acmISBN{979-8-4007-2259-2/2026/08}
\acmDOI{10.1145/3770855.3818994}

\usepackage{algorithm}
\usepackage{algorithmic}
\usepackage{multirow}
\usepackage{microtype}
\usepackage{float}
\usepackage{stfloats}
\usepackage{subcaption}
\usepackage{colortbl}

\newcommand\kddavailabilityurl{https://doi.org/10.5281/zenodo.20261896}

\newcommand{\grayrow}{\rowcolor[gray]{.90}}

\begin{document}

\title[Cosmo3DFlow: Wavelet Flow Matching for the Early Universe]{Cosmo3DFlow: Wavelet Flow Matching for Spatial-to-Spectral Compression in Reconstructing the Early Universe}

\author{Md. Khairul Islam}
\email{mi3se@virginia.edu}
\orcid{0000-0003-2894-8584}
\affiliation{%
  \department{Department of Computer Science}
  \institution{University of Virginia}
  \city{Charlottesville}
  \state{VA}
  \country{USA}
}

\author{Zeyu Xia}
\email{zeyu.xia@virginia.edu}
\orcid{0000-0003-0234-5857}
\affiliation{%
  \department{Department of Computer Science}
  \institution{University of Virginia}
  \city{Charlottesville}
  \state{VA}
  \country{USA}
}

\author{Ryan Goudjil}
\email{kqs9vd@virginia.edu}
\orcid{0009-0008-1024-7708}
\affiliation{%
  \department{Department of Computer Science}
  \institution{University of Virginia}
  \city{Charlottesville}
  \state{VA}
  \country{USA}
}

\author{Jialu Wang}
\email{walter.wang@utexas.edu}
\orcid{0009-0007-0014-2657}
\affiliation{%
  \department{Department of Statistics and Data Sciences}
  \institution{The University of Texas at Austin}
  \city{Austin}
  \state{TX}
  \country{USA}
}

\author{Arya Farahi}
\email{arya.farahi@austin.utexas.edu}
\orcid{0000-0003-0777-4618}
\affiliation{%
  \department{Department of Statistics and Data Sciences}
  \institution{The University of Texas at Austin}
  \city{Austin}
  \state{TX}
  \country{USA}
}

\author{Judy Fox}
\authornote{Corresponding author.}
\email{ckw9mp@virginia.edu}
\orcid{0000-0001-8198-4117}
\affiliation{
  \department[1]{School of Data Science}
  \department[2]{Department of Computer Science}
  \institution{University of Virginia}
  \city{Charlottesville}
  \state{VA}
  \country{USA}
}

\renewcommand{\shortauthors}{Md. Khairul Islam, Zeyu Xia, Ryan Goudjil, Jialu Wang, Arya Farahi, and Judy Fox}

\begin{abstract}
  Reconstructing the early universe from the evolved present-day universe is a challenging and computationally demanding problem in modern astrophysics. We devise a novel generative framework, Cosmo3DFlow, designed to address dimensionality and sparsity, the critical bottlenecks inherent in current state-of-the-art methods for cosmological inference. By integrating 3D Discrete Wavelet Transform (DWT) with flow matching, we effectively represent high-dimensional cosmological structures. The Wavelet Transform addresses the ``void problem'' by translating spatial emptiness into spectral sparsity. It decouples high-frequency details from low-frequency structures, and wavelet-space velocity fields facilitate stable ordinary differential equation (ODE) solvers with large step sizes. Using large-scale cosmological $N$-body simulations at $128^3$ resolution, we achieve up to $46\times$ faster sampling than diffusion models. Our results enable initial conditions to be sampled in seconds, compared to minutes for previous methods.
\end{abstract}



\maketitle
\ifdefempty{\kddavailabilityurl}{}{
  \begingroup\small\noindent\raggedright\textbf{Resource Availability:}\\
  The source code of this paper has been made publicly available at \url{\kddavailabilityurl} and on GitHub at \url{https://github.com/UVA-MLSys/Cosmo3DFlow}.
  \endgroup
}

\section{Introduction}
Understanding the origins of the universe relies on mapping the complex, non-linear structures observed today back to their initial conditions~\cite{hutererCourseCosmologyTheory2023}. This is a \textit{high-dimensional inverse problem} of staggering complexity, often described as trying to un-mix a fluid to determine its initial state. In the context of cosmology, the ``fluid'' is the distribution of dark matter and baryonic matter that forms galaxies, stars, and gas. The ``mixing'' is driven by gravitational instability as the universe evolves.

\begin{figure}[!htbp]
  \centering
  \includegraphics[width=.8\linewidth]{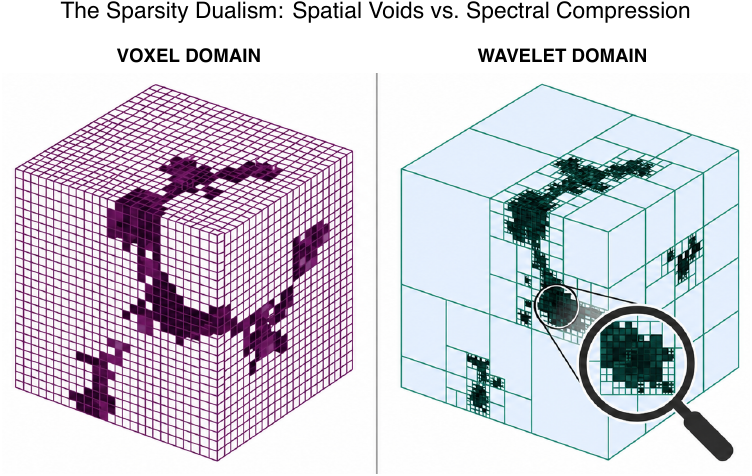}
  \caption{Cosmic Web data representations. Left: Voxel grids allocate memory uniformly, wasting computation on empty voids. Right: Wavelets concentrate informative coefficients on filaments and halos, leaving void regions as near-zero coefficients.}
  \Description{A side-by-side diagram titled ``The Sparsity Dualism'' compares a sparse 3D structure plotted on a uniform voxel grid against its representation on a hierarchical wavelet grid, where a magnifying glass highlights the adaptive, localized resolution used for spectral compression.}%
  \label{fig:sparsity}
\end{figure}

\begin{figure*}[!t]
  \centering
  \includegraphics[width=\textwidth]{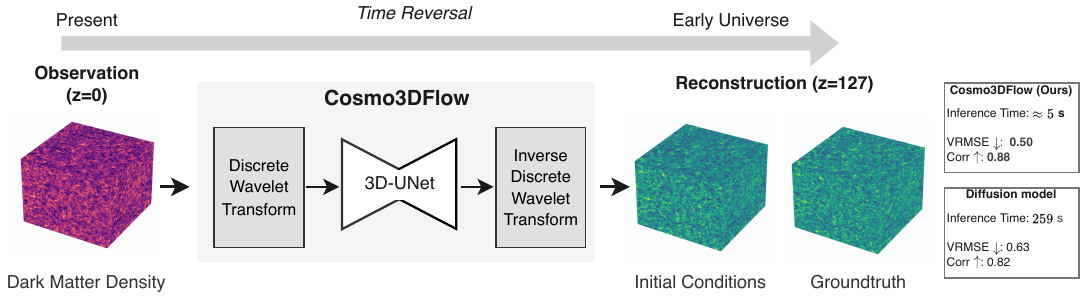}
  \caption{\textbf{Cosmo3DFlow accelerates high-dimensional inverse problems with $46\times$ faster sampling.} It reconstructs samples in ${\approx}5$\,s (vs.\ $259$\,s for diffusion), preserving fine-scale structure through wavelet sparsity while achieving superior variance-normalized root mean squared error (VRMSE) and correlation metrics.}
  \Description{Comparison of cosmic density field reconstruction showing the evolved universe at redshift 0, Cosmo3DFlow reconstruction at redshift 127, with a zoom-in panel highlighting fine-scale structure preservation.}%
  \label{fig:teaser}
\end{figure*}

Analysis of $N$-body simulations indicates a stark disconnect between the volume a structure occupies and the mass it contains. Voids are the dominant topological feature by volume: data derived from dark matter simulations indicate that voids occupy approximately 63.7\% of the total volume~\cite{metukiGalaxyPropertiesCosmic2015}. However, despite this volumetric dominance, they effectively represent ``empty'' computational space, containing only 16.2\% of the dark matter mass. In contrast, filaments and halos---the regions of primary astrophysical interest where galaxies form---are sparse but mass-dense. Filaments occupy roughly 4.3\% of the volume but house 33.0\% of the mass, while halos are even more extreme, representing a tiny volumetric fraction yet containing significant mass concentrations.

The cosmology inference challenge is formidable. For a simulation on a $128^3$ grid (approximately 2.1 million voxels), a 63.7\% void fraction implies that over 1.3 million voxels are dedicated to describing regions where the density contrast is low and the gradient is near zero. A typical full $N$-body simulation (forward model) mapping initial conditions to observations requires numerically simulating gravitational dynamics via $N$-body methods, where traditional Markov Chain Monte Carlo (MCMC) approaches require thousands of forward simulations per sample. In standard Convolutional Neural Networks (CNNs) or diffusion models operating in voxel space, every one of these 1.3 million void voxels requires the same number of floating-point operations (FLOPs) as the dense, complex voxels inside a galaxy halo. This is the ``uniform computation cost'' mismatch of the voxel grid: resources are allocated based on volume, not information density. The voxel representation forces a ``worst-case scenario'' resolution across the entire domain as the solver must be fine enough everywhere to handle the densest regions, leading to massive redundancy in the sparse regions (see the left panel of Figure~\ref{fig:sparsity}). This mismatch between the uniform sampling of the Eulerian grid and the heterogeneous clustering of matter is the fundamental physical driver for adopting multi-scale representations such as hierarchical ball trees~\cite{zhdanovErwinTreebasedHierarchical2025}, 3D point clouds~\cite{chatterjeeCosmologyPointClouds2025}, and wavelets.

Recent advances have opened directions for high-dimensional problems. Score-based diffusion has been applied to this problem~\cite{leginPosteriorSamplingInitial2024} but remains computationally expensive. Two recent developments in generative modeling suggest potential improvements: (1) \textit{Flow matching} formulates generation as solving a deterministic ordinary differential equation (ODE) rather than a stochastic process, typically enabling faster and more stable sampling with fewer integration steps~\cite{lipmanFlowMatchingGenerative2023}. (2) \textit{Wavelet-based generative models} have demonstrated that operating in the wavelet coefficient domain rather than pixel space can provide substantial computational savings while preserving the high-frequency details crucial for many applications~\cite{friedrichWDM3DWavelet2025, mebratuWaveletFlowExtragalactic2026}. We hypothesize that combining these two innovations is particularly well-suited for cosmological inference due to the hierarchical nature of structure formation~\cite{whiteGalaxyFormationHierarchical1991}.

We introduce \textbf{Cosmo3DFlow}, a framework combining flow matching with wavelet-space modeling for high-dimensional inverse problems (Figure~\ref{fig:teaser}). Cosmo3DFlow achieves order of magnitude speedups over diffusion baselines while matching or improving reconstruction quality. \textbf{Our main contributions include:}

\begin{itemize}
  \item We combine the \textbf{3D Discrete Wavelet Transform} with \textbf{flow matching} for cosmological inverse problems. The DWT losslessly reshapes the field into eight wavelet subbands at half spatial resolution, enabling $2\times$ larger training batches and lower per-step inference cost.
  \item We introduce wavelet-aware scale-specific conditioning that exploits the multi-resolution structure of DWT coefficients, and power spectrum regularization that balances reconstruction error across wavenumbers during training.
  \item On \textbf{Quijote} $128^3$ simulations, Cosmo3DFlow achieves $\approx 46\times$ faster sampling than diffusion models while matching or improving reconstruction fidelity, measured by lower variance-normalized root mean squared error (VRMSE), higher correlation, and accurate power spectrum recovery.
\end{itemize}
The methodological contribution of Cosmo3DFlow lies in the exact mathematical coupling of Optimal Transport (Flow Matching) with discrete spectral bijections. By replacing lossy neural autoencoders with the Discrete Wavelet Transform (DWT), it achieves a lossless spatial-to-spectral compression---an exact reshape under which sparse spatial structure (voids) collapses into spectrally sparse coefficients. This methodology structurally smooths the learned optimal transport vector field---bounding its Lipschitz constant to enable ultra-fast, numerically stable ODE integration. Ultimately, Cosmo3DFlow establishes a highly efficient methodology for scaling sparse high-dimensional generative AI models for the physical sciences.
\section{Related Work}

\noindent \textbf{The Cosmological Inference Challenge.} The standard model of cosmology ($\Lambda$CDM) posits that the cosmic web evolved from initial Gaussian Random Field fluctuations~\cite{ModernCosmology2021}. Under gravitational instability, peaks of this Gaussian field collapse into virialized halos while underdense regions expand to form voids~\cite{bardeenStatisticsPeaksGaussian1986}. Reconstructing the initial density field, denoted as $\mathbf{x}$, from the evolved observation $\mathbf{y}$ can be framed as a Bayesian inference task. We sample from the posterior distribution $p(\mathbf{x}|\mathbf{y}) = p(\mathbf{y}|\mathbf{x})p(\mathbf{x}) / p(\mathbf{y})$. Here, the likelihood $p(\mathbf{y}|\mathbf{x})$ encapsulates the forward model of gravitational evolution (typically an $N$-body simulation), and the prior $p(\mathbf{x})$ is the Gaussian prior of the initial conditions.

The challenge is twofold: \textbf{Dimensionality.} As resolutions increase, the dimensionality explodes. \textbf{Computational Cost.} Traditional statistical approaches like Bayesian Origin Reconstruction from Galaxies (BORG)~\cite{jascheBayesianPhysicalReconstruction2013, jaschePhysicalBayesianModelling2019} utilize Hamiltonian Monte Carlo to explore this high-dimensional landscape. However, this involves running the expensive forward simulation and can take thousands of CPU hours.

\noindent \textbf{Scientific significance.} Mapping present-day non-linear structures back to their initial conditions is a fundamental inverse problem for probing early-universe physics (e.g., inflation, neutrino masses). Discrepancies in modeling probes like galaxy clustering and the cosmic microwave background can reveal new physics. High-fidelity suites like Quijote~\cite{villaescusa-navarroQuijoteSimulations2020} provide the controlled environment needed to validate these methods.

\noindent \textbf{The Generative AI Paradigm Shift.} Recent advances in deep generative modeling offer a transformative alternative: training a neural network to approximate the posterior directly. Modi et al.~\cite{modiCosmicRIMReconstructingEarly2021} predicted point estimates of initial conditions via recurrent inference machines coupled with a differentiable $N$-body simulator; Shallue and Eisenstein~\cite{shallueReconstructingCosmologicalInitial2023} explored convolutional neural networks for deterministic reconstruction, and Riveros et al.~\cite{riverosConditionalDiffusionFlowModels2025} applied diffusion to generate 3D density fields of modified $f(R)$ gravity. Legin et al.~\cite{leginPosteriorSamplingInitial2024} applied score-based diffusion~\cite{hoDenoisingDiffusionProbabilistic2020, songImprovedTechniquesTraining2020} to cosmological initial conditions inference, though stochastic differential equation (SDE) sampling in high-dimensional pixel space incurs substantial cost. Latent diffusion models have also been used for spatiotemporal physics emulations~\cite{rozetLostLatentSpace2026} and simulation super-resolution~\cite{mishraCosmoFOLDFastGeneration2026}.

\noindent \textbf{Flow Matching.} Flow matching learns velocity fields via simple regression, offering more stable training and deterministic ODE sampling with fewer integration steps than stochastic SDEs~\cite{lipmanFlowMatchingGenerative2023}. CosmoFlow~\cite{kannanCosmoFlowScaleAwareRepresentation2025} demonstrated flow matching for cosmological representation learning but operated at lower resolution (2D, $256 \times 256$) without sampling full 3D volumes. FlowLensing~\cite{sayedFlowLensingSimulatingGravitational2025} applied flow matching to 2D gravitational lensing but operated in pixel space without addressing 3D volumetric inverse problems.

\noindent \textbf{Wavelet-Based Generative Models.} Recent works~\cite{phungWaveletDiffusionModels2023, friedrichWDM3DWavelet2025, sigilloLatentWaveletDiffusion2026} demonstrated that diffusion models in wavelet space enable efficient high-resolution generation. Phung et al.~\cite{phungWaveletDiffusionModels2023} showed that wavelet diffusion models can match or exceed the quality of latent diffusion models while avoiding the need to train auxiliary autoencoders. Friedrich et al.~\cite{friedrichWDM3DWavelet2025} adapted discrete wavelet diffusion for high-resolution 3D medical imaging. Mebratu and Wu~\cite{mebratuWaveletFlowExtragalactic2026} used Wavelet Flow to model the field-level probability distributions of extragalactic foregrounds in cosmic observations. Valogiannis et al.~\cite{valogiannisUnveilingLargescaleNature2024} applied Wavelet Scattering Transform to constrain the nature of gravity using 3D $N$-body simulations.

\section{Methodology}

\subsection{Problem Formulation}

We formalize our goal as a conditional generation problem: given an observation $\mathbf{y} \in \mathbb{R}^{N^3}$ (evolved state), sample from the posterior $p(\mathbf{x}|\mathbf{y})$ over latent states $\mathbf{x} \in \mathbb{R}^{N^3}$ (initial conditions). The forward process $\mathbf{y} = f(\mathbf{x})$ is nonlinear and expensive to evaluate (requiring numerical simulation), making traditional MCMC intractable. Our goal is a generative model that produces high-quality samples conditioned on $\mathbf{y}$.

A natural response to this sparsity is to employ a sparse generative model that omits empty regions, as is common in natural image generation. Such an approach is ill-suited here; cosmological voids are not truly empty: they exert weak gravitational influence on surrounding structure and encode information about large-scale modes, and discarding them would remove a physically meaningful signal. We therefore retain a dense formulation, but cast it in the wavelet domain, where voids collapse into a small number of low-magnitude coefficients and are compactly represented rather than excised.

\subsection{Flow Matching Framework}

Flow matching learns continuous normalizing flows by regressing velocity fields that transport samples from a simple prior $p_0$ to the target data distribution $p_1$~\cite{lipmanFlowMatchingGenerative2023}. Given samples from the prior $\mathbf{x}_0 \sim p_0$ and target $\mathbf{x}_1 \sim p_1$, we define an interpolation path:
\begin{equation}
  \mathbf{x}_t = \mu_t(\mathbf{x}_0, \mathbf{x}_1) = t \mathbf{x}_1 + (1-t) \mathbf{x}_0, \quad t \in [0, 1]
\end{equation}
This linear interpolation, known as rectified flow~\cite{liuFlowStraightFast2022}, yields a constant target velocity field:
\begin{equation}
  \mathbf{u}_t = \frac{d\mathbf{x}_t}{dt} = \mathbf{x}_1 - \mathbf{x}_0
\end{equation}

For conditional generation with observation $\mathbf{y}$, we train a velocity network $\mathbf{v}_\theta(\mathbf{x}_t, t, \mathbf{y})$ to approximate this target velocity by minimizing the flow matching objective:
\begin{equation}
  \mathcal{L}_{\text{flow}} = \mathbb{E}_{t \sim \mathcal{U}(0,1), \mathbf{x}_0 \sim p_0, \mathbf{x}_1 \sim p_1} \left[ \| \mathbf{v}_\theta(\mathbf{x}_t, t, \mathbf{y}) - \mathbf{u}_t \|_2^2 \right]
\end{equation}

At inference time, we sample initial noise $\mathbf{x}_0 \sim \mathcal{N}(\mathbf{0}, \mathbf{I})$ and integrate the learned velocity field to obtain the final sample:
\begin{equation}
  \mathbf{x}_1 = \mathbf{x}_0 + \int_0^1 \mathbf{v}_\theta(\mathbf{x}_t, t, \mathbf{y}) \, dt
\end{equation}
where the integral accumulates the velocity contributions along the flow trajectory, transporting the initial noise $\mathbf{x}_0$ to the target distribution. We approximate this integral using a numerical ODE solver with $K$ discrete steps. By a standard Euler stability argument, the maximum stable step size scales inversely with the Lipschitz constant $\mathrm{Lip}(\mathbf{v}_\theta)$ of the learned velocity field, so a smaller $\mathrm{Lip}(\mathbf{v}_\theta)$ admits fewer integration steps (\textbf{Proposition~\ref{prop:ode_lipschitz_bounds}}). The deterministic nature of ODE integration enables more stable and efficient generation.

\subsection{Wavelet Flow Matching}

\noindent \textbf{Wavelet Domain Transformation.} The Discrete Wavelet Transform (DWT) consists of separable convolutions with low-pass ($L$) and high-pass ($H$) filters followed by downsampling by a factor of 2 along each spatial dimension. For a 3D volume $\mathbf{y} \in \mathbb{R}^{N^3}$, this decomposition yields eight wavelet coefficient tensors:
\begin{equation}
  \text{DWT}(\mathbf{y}) = (\mathbf{c}_{LLL}, \mathbf{c}_{LLH}, \mathbf{c}_{LHL}, \mathbf{c}_{LHH}, \mathbf{c}_{HLL}, \mathbf{c}_{HLH}, \mathbf{c}_{HHL}, \mathbf{c}_{HHH})
\end{equation}

where each coefficient tensor $\mathbf{c}_{*}$ has dimensions $(N/2)^3$. We use the Haar wavelet basis, which applies box-averaging for the low-pass filter and pairwise differencing for the high-pass filter along each axis. To verify this choice, we tested smoother alternatives (Daubechies-2, Symlet-2, Biorthogonal-2.2) from the \texttt{PyWavelets} library on a $32^3$ grid. Haar remains optimal: it achieves the fastest per-epoch training ($5.32$\,s vs.\ $5.54$\,s for the next best), the lowest VRMSE ($0.283$ vs.\ $0.297$), and the best physical scores (e.g., cross-correlation $0.975$ vs.\ $0.970$). Across wavelet types we do not observe substantial differences in accuracy, supporting Haar as a fast and accurate default.

With all coefficients retained and consistent periodic boundary handling, this transform is exactly invertible (\textbf{Proposition~\ref{prop:dwt_exactness}}). The approximation coefficient $\mathbf{c}_{LLL}$ captures low-frequency content representing the large-scale skeleton of the cosmic web (filaments traced at coarse resolution and the interiors of voids), while the remaining seven detail coefficients encode high-frequency information aligned with halos and fine filamentary structure along different directional orientations. We stack these coefficients along the channel dimension to form a unified representation:
\begin{equation}
  \tilde{\mathbf{y}} = \text{Stack}(\mathbf{c}_{LLL} / \sqrt{8}, \mathbf{c}_{LLH}, \ldots, \mathbf{c}_{HHH}) \in \mathbb{R}^{8 \times (N/2)^3}
  \label{eq:wavelet_stack}
\end{equation}
where we normalize $\mathbf{c}_{LLL}$ by $\sqrt{8}$ to match the dynamic range across coefficients. The original field can be exactly recovered via the Inverse Discrete Wavelet Transform (IDWT): $\mathbf{y} = \text{IDWT}(\tilde{\mathbf{y}})$, where, by convention throughout this paper, $\text{IDWT}$ first multiplies the $\mathbf{c}_{LLL}$ channel by $\sqrt{8}$ to undo this range-matching normalization before applying the standard inverse Haar transform.

The wavelet representation offers three key advantages for cosmological inference:
\begin{enumerate}
  \item \textbf{Sparse representations:} Wavelet coefficients of cosmological density fields exhibit sparse, structured distributions that are more amenable to generative modeling than dense voxel representations.
  \item \textbf{Per-layer spatial-axis reduction:} Halving each spatial axis lets 3D convolutions after the input layer operate on volumes that are $8\times$ smaller while preserving complete information (Proposition~\ref{prop:dwt_exactness}), reducing memory and computational cost.
  \item \textbf{Physical alignment:} The multi-resolution decomposition mirrors the hierarchical nature of cosmological structure formation, where large-scale density fluctuations are established early during inflation and small-scale structures develop through subsequent gravitational collapse.
\end{enumerate}

\noindent \textbf{Wavelet-Space Flow Matching.} We formulate the flow matching objective entirely in wavelet space. Let $\tilde{\mathbf{x}}_0 = \text{DWT}(\mathbf{x}_0)$ and $\tilde{\mathbf{x}}_1 = \text{DWT}(\mathbf{x}_1)$ denote the wavelet-transformed noise and target fields, respectively. The wavelet-space loss becomes:
\begin{equation}
  \tilde{\mathcal{L}}_{\text{flow}} = \mathbb{E}_{t, \tilde{\mathbf{x}}_0, \tilde{\mathbf{x}}_1} \left[ \| \mathbf{v}_\theta(\tilde{\mathbf{x}}_t, t, \tilde{\mathbf{y}}) - (\tilde{\mathbf{x}}_1 - \tilde{\mathbf{x}}_0) \|_2^2 \right]
\end{equation}
where $\tilde{\mathbf{x}}_t = t \tilde{\mathbf{x}}_1 + (1-t) \tilde{\mathbf{x}}_0$ is the interpolated state in wavelet space, and $\tilde{\mathbf{y}} = \text{DWT}(\mathbf{y})$ is the wavelet-transformed conditioning observation.

At inference time, we sample initial Gaussian noise $\mathbf{x}_0 \sim \mathcal{N}(\mathbf{0}, \mathbf{I})$ in voxel space, transform to the wavelet domain, integrate the velocity field, and apply the inverse wavelet transform to recover the final density field:
\begin{equation}
  \hat{\mathbf{x}} = \text{IDWT}\left( \tilde{\mathbf{x}}_0 + \int_0^1 \mathbf{v}_\theta(\tilde{\mathbf{x}}_t, t, \tilde{\mathbf{y}}) \, dt \right)
\end{equation}
In practice, we approximate this integral with a numerical ODE solver (e.g.,~Euler) over $K$ discrete steps (Algorithm~\ref{alg:inference}).

\noindent \textbf{Power Spectrum Regularization.} To balance reconstruction error across wavenumbers, we incorporate a log-spectral consistency loss on the velocity field. The 3D isotropic power spectrum $P(k)$ characterizes the variance of a field as a function of wavenumber $k = |\mathbf{k}|$, which is computed as:
\begin{equation}
  P(k_i) = \frac{1}{N_{\text{modes}}(k_i)} \sum_{|\mathbf{k}| \in \text{bin}_i} |\tilde{f}(\mathbf{k})|^2
\end{equation}
where $\tilde{f}(\mathbf{k})$ denotes the Fourier transform of the field and $N_{\text{modes}}(k_i)$ counts modes in the $i$-th radial bin.

The power spectrum loss penalizes discrepancies in log-space to treat relative errors uniformly across the dynamic range:
\begin{equation}
  \mathcal{L}_{\text{PS}} = \|\log P_{\text{pred}}(k) - \log P_{\text{target}}(k)\|_2^2
\end{equation}
This regularization is applied to the predicted and target velocities after they are transformed back to voxel space via the inverse DWT\@. The combined training objective becomes:
\begin{equation}
  \mathcal{L}_{\text{total}} = \tilde{\mathcal{L}}_{\text{flow}} + \lambda_{\text{PS}} \mathcal{L}_{\text{PS}}
\end{equation}
where $\lambda_{\text{PS}}$ is a hyperparameter controlling the strength of the power spectrum regularization. We keep the value at $0.01$.

\subsection{Theoretical Analysis}
We present two theoretical foundations that underpin our work: the existence of a stable lower bound on the ODE step size, and the exactness of the spatial-to-spectral transform.

\begin{proposition}[\textbf{ODE Stable Step-Size Lower Bound}]\label{prop:ode_lipschitz_bounds}
Let the learned velocity field $\mathbf{v}_\theta(\cdot, t)$ be Lipschitz in its state argument with constant $\mathrm{Lip}(\mathbf{v}_\theta)$. Then the explicit Euler scheme $\mathbf{z}_{k+1} = \mathbf{z}_k + \Delta t\, \mathbf{v}_\theta(\mathbf{z}_k, t_k)$ admits a stable step size satisfying $\Delta t \le c/\mathrm{Lip}(\mathbf{v}_\theta)$, where $c > 0$ depends only on the chosen stability region~\cite{SolvingOrdinaryDifferential1993}. Consequently, the maximum stable step size scales as
\begin{equation}
\Delta t_{\max} \;\propto\; \frac{1}{\mathrm{Lip}(\mathbf{v}_\theta)},
\end{equation}
so a velocity field with a smaller Lipschitz constant admits a larger guaranteed stable step and therefore a smaller discrete step count $K = \lceil 1/\Delta t \rceil$ for accurate trajectory integration.
\end{proposition}

\begin{proposition}[\textbf{Exactness of Orthogonal DWT}]\label{prop:dwt_exactness}
Let $\mathbf{x} \in \mathbb{R}^N$ be a discrete, finite-dimensional spatial input, and let $\mathbf{W} \in \mathbb{R}^{N \times N}$ denote the orthogonal discrete wavelet transform matrix. Because $\mathbf{W}^T \mathbf{W} = \mathbf{I}_N$, the forward mapping $\mathbf{z} = \mathbf{W}\mathbf{x}$ and its inverse $\hat{\mathbf{x}} = \mathbf{W}^T \mathbf{z}$ form an exact linear bijection, satisfying
\begin{equation}
\|\mathbf{x} - \mathbf{W}^T(\mathbf{W}\mathbf{x})\|_2 = 0
\end{equation}
up to machine precision. The range-matching normalization of $\mathbf{c}_{LLL}$ by $\sqrt{8}$ in Eq.~\eqref{eq:wavelet_stack} corresponds to left-multiplying $\mathbf{W}$ by an invertible diagonal matrix $\mathbf{D}$, so the full pipeline operator $\mathbf{D}\mathbf{W}$ remains an exact bijection with inverse $\mathbf{W}^T \mathbf{D}^{-1}$.
\end{proposition}

The spatial-to-wavelet mapping is therefore strictly lossless, limited only by machine epsilon, so any approximation error in our pipeline stems entirely from sources outside this transform.


\subsection{Network Architecture}
\noindent \textbf{3D U-Net Velocity Network.} Our velocity network $\mathbf{v}_\theta$ is a 3D U-Net (Figure~\ref{fig:3dunet}) that takes 16 input channels (the concatenation of the noisy wavelet coefficients $\tilde{\mathbf{x}}_t$ and the conditioning observation $\tilde{\mathbf{y}}$, each in $\mathbb{R}^{8 \times (N/2)^3}$) and predicts an 8-channel velocity in wavelet space. The encoder downsamples to a fixed $8 \times 8 \times 8$ bottleneck (2, 3, and 4 stages for $32^3$, $64^3$, and $128^3$ inputs, respectively), and the decoder mirrors it with skip connections. Each level uses $N_{\text{res}} = 2$ BigGAN-style residual blocks conditioned on a Gaussian Fourier time embedding~\cite{tancikFourierFeaturesLet2020}, with base feature dimension $n_f = 32$. The full architecture, including residual block equations, time embedding, and forward-pass pseudocode, is detailed in Appendix~\ref{sec:appendix_unet}.

To exploit the structure of wavelet representations, we introduce two wavelet-aware architectural modifications.

\noindent \textbf{Scale-Specific Conditioning.} Rather than treating all wavelet bands uniformly, we inject scale-specific information at each resolution level of the U-Net. At resolution level $i$, we project the 8-channel wavelet observation through a $1 \times 1 \times 1$ convolution:
\begin{equation}
  \mathbf{z}_i = \text{Conv}_{1 \times 1 \times 1}(\tilde{\mathbf{y}}) \in \mathbb{R}^{n_f \cdot m_i \times N_i^3}
\end{equation}
followed by adaptive average pooling to match the spatial dimensions. This conditioning is added to the hidden state after each residual block, providing direct access to scale-relevant information at each resolution.

\noindent \textbf{Cross-Scale Skip Connections.} To facilitate information flow between different wavelet scales during generation, we introduce additional skip connections that bridge encoder features directly to corresponding decoder levels through learned $1 \times 1 \times 1$ projections:
\begin{equation}
  \mathbf{h}_{\text{dec}}^{(i)} = \mathbf{h}_{\text{dec}}^{(i)} + \text{Conv}_{1 \times 1 \times 1}(\mathbf{h}_{\text{enc}}^{(i)})
\end{equation}
These connections allow the model to maintain coherent scale relationships throughout the generation process.

\subsection{Training and Inference}

The complete training and inference procedures are detailed in Algorithm~\ref{alg:training} and Algorithm~\ref{alg:inference}, respectively.

\begin{algorithm}[!htbp]
  \caption{\textit{Cosmo3DFlow Training}}\label{alg:training}
  \setlength{\fboxsep}{1pt}%
  \begin{algorithmic}[1]
    \STATE \textbf{Input:} Dataset $\mathcal{D} = \{(\mathbf{x}_i, \mathbf{y}_i)\}_{i=1}^{M}$, velocity model $\mathbf{v}_\theta$, power-spectrum weight $\lambda_{\text{PS}}$
    \STATE \textbf{Output:} Trained parameters $\theta^*$

    \STATE Initialize $\mathbf{v}_\theta$ and EMA parameters with decay $0.99$

    \FOR{epoch $e = 1, \dots, E_{\max}$}
      \FOR{each batch $(\mathbf{x}, \mathbf{y})$}
        \STATE \textit{\textbf{Stage 1: Wavelet encoding}}
        \STATE Sample $t \sim \mathcal{U}(0,1)$ and $\mathbf{x}_0 \sim \mathcal{N}(\mathbf{0}, \mathbf{I})$
        \STATE \colorbox{gray!20}{\parbox{0.95\linewidth}{Wavelet transform: $\tilde{\mathbf{x}}_0 \!\gets\! \text{DWT}(\mathbf{x}_0)$, $\tilde{\mathbf{x}}_1 \!\gets\! \text{DWT}(\mathbf{x})$, $\tilde{\mathbf{y}} \!\gets\! \text{DWT}(\mathbf{y})$}}
        \STATE Interpolate: $\tilde{\mathbf{x}}_t \gets t\,\tilde{\mathbf{x}}_1 + (1-t)\,\tilde{\mathbf{x}}_0$
        \STATE \textit{\textbf{Stage 2: Velocity prediction}}
        \STATE \colorbox{gray!20}{\parbox{0.95\linewidth}{Predict velocity via Algorithm~\ref{alg:unet}: $\mathbf{v}_t \gets \mathbf{v}_\theta(\tilde{\mathbf{x}}_t, t, \tilde{\mathbf{y}})$}}
        \STATE \textit{\textbf{Stage 3: Loss computation and update}}
        \STATE Flow-matching loss: $\tilde{\mathcal{L}}_{\text{flow}} \gets \|\mathbf{v}_t - (\tilde{\mathbf{x}}_1 - \tilde{\mathbf{x}}_0)\|_2^2$
        \STATE \colorbox{gray!20}{\parbox{0.95\linewidth}{Predicted power spectrum:\\ $P_{\text{pred}} \gets \text{PowerSpec}(\text{IDWT}(\mathbf{v}_t))$}}
        \STATE \colorbox{gray!20}{\parbox{0.95\linewidth}{Target power spectrum:\\ $P_{\text{target}} \gets \text{PowerSpec}(\text{IDWT}(\tilde{\mathbf{x}}_1 - \tilde{\mathbf{x}}_0))$}}
        \STATE {\parbox{0.95\linewidth}{Power-spectrum loss:\\ $\mathcal{L}_{\text{PS}} \gets \|\log P_{\text{pred}} - \log P_{\text{target}}\|_2^2$}}
        \STATE Total loss: $\mathcal{L}_{\text{total}} \gets \tilde{\mathcal{L}}_{\text{flow}} + \lambda_{\text{PS}} \mathcal{L}_{\text{PS}}$; update $\theta$ by gradient descent; update EMA
      \ENDFOR
    \ENDFOR

    \STATE \textbf{Return:} EMA-averaged $\theta^*$
  \end{algorithmic}
  \parbox{\columnwidth}{\small\textit{Note:} \colorbox{gray!20}{gray} marks our modifications.}
\end{algorithm}

\begin{algorithm}[!htbp]
  \caption{\textit{Cosmo3DFlow Inference: Sample Initial Conditions}}\label{alg:inference}
  \setlength{\fboxsep}{1pt}%
  \begin{algorithmic}[1]
    \STATE \textbf{Input:} Observed $\mathbf{y}$, trained EMA model $\mathbf{v}_{\theta^*}$, ODE steps $K$
    \STATE \textbf{Output:} Initial condition $\hat{\mathbf{x}}$

    \STATE \textit{\textbf{Stage 1: Wavelet encoding}}
    \STATE \colorbox{gray!20}{\parbox{0.95\linewidth}{Compute $\tilde{\mathbf{y}} \gets \text{DWT}(\mathbf{y})$}}
    \STATE \colorbox{gray!20}{\parbox{0.95\linewidth}{Sample $\mathbf{x}_0 \sim \mathcal{N}(\mathbf{0}, \mathbf{I})$ in voxel space and set $\tilde{\mathbf{x}}_0 \gets \text{DWT}(\mathbf{x}_0)$}}

    \STATE \textit{\textbf{Stage 2: ODE integration}}
    \FOR{$n = 1$ to $K$}
      \STATE Set $t \gets (n-1)/K$
      \STATE Predict velocity: $\mathbf{v} \gets \mathbf{v}_{\theta^*}(\tilde{\mathbf{x}}_{n-1}, t, \tilde{\mathbf{y}})$
      \STATE Euler step: $\tilde{\mathbf{x}}_n \gets \tilde{\mathbf{x}}_{n-1} + (1/K)\,\mathbf{v}$
    \ENDFOR

    \STATE \textit{\textbf{Stage 3: Decoding}}
    \STATE \colorbox{gray!20}{\parbox{0.95\linewidth}{Recover $\hat{\mathbf{x}} \gets \text{IDWT}(\tilde{\mathbf{x}}_K)$}}
    \STATE \textbf{Return:} $\hat{\mathbf{x}}$
  \end{algorithmic}
\end{algorithm}

\noindent \textbf{Training.}
We use AdamW optimization with learning rate $\eta = 10^{-4}$ and a ReduceLROnPlateau scheduler with patience 5 and factor 0.5. Gradient clipping with maximum norm 1.0 ensures stable training. Data augmentation via random 3D rotations and reflections improves generalization. Each model is trained for 100 epochs, and the best model by validation loss is selected for testing. We use batch sizes 16, 8, and 4 for the $32^3$, $64^3$, and $128^3$ resolutions, respectively. All training and inference use the NVIDIA A100 GPU with 80\,GB of memory.

\noindent \textbf{Inference.} At test time, we sample Gaussian noise in voxel space, transform to the wavelet domain, integrate the learned velocity field using the Euler method with $K$ steps, and recover the final density field via the IDWT\@. The wavelet-space formulation enables accurate sampling with far fewer steps than diffusion-based approaches, consistent with the Euler stability lower bound on the ODE step size.

\section{Experiments}
\begin{figure*}[!tb]
  \centering
  \includegraphics[width=.8\textwidth]{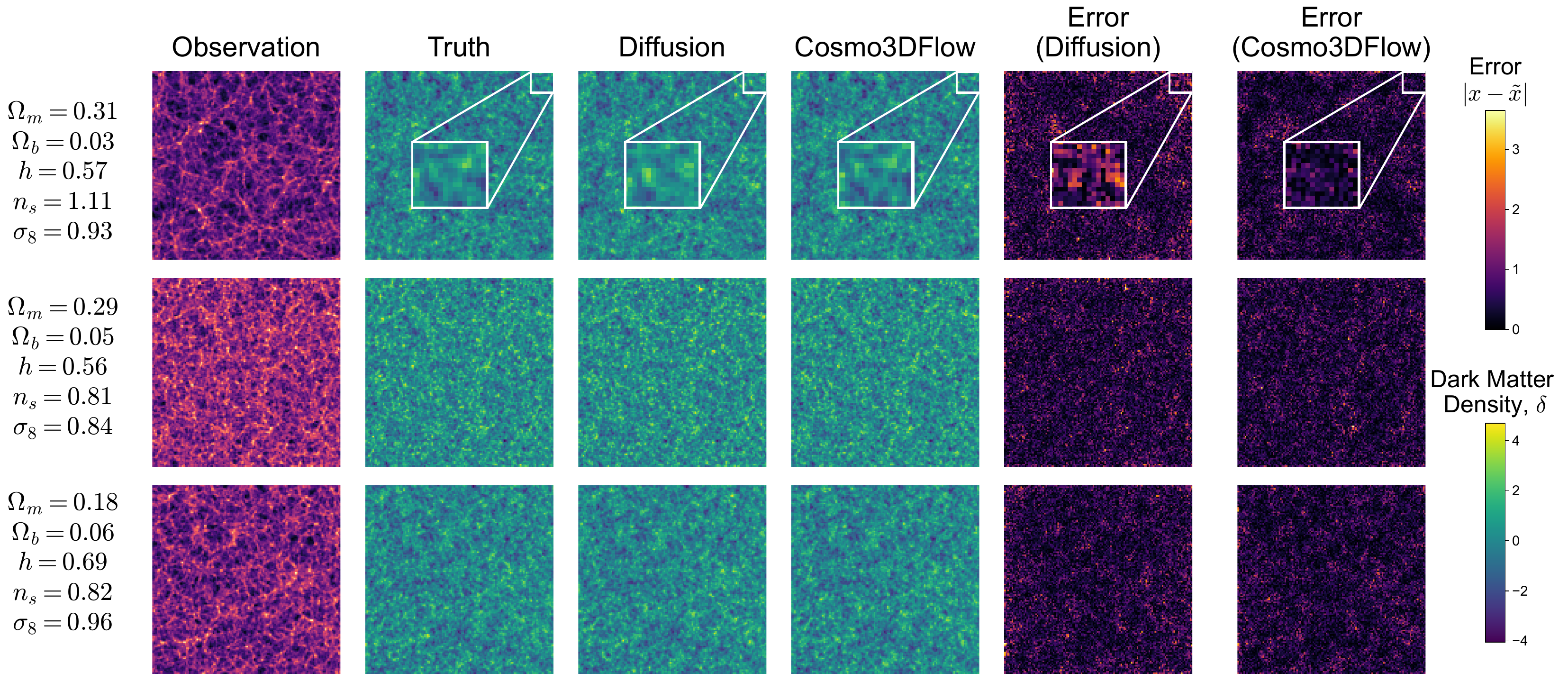}
  \caption{\textbf{Qualitative comparison of reconstruction quality.} We show 2D slices of 3D density fields from the Standard Latin Hypercube, reconstructing initial conditions from observations. Compared to diffusion, Cosmo3DFlow preserves sharp, small-scale structures (highlighted boxes) without blurring artifacts, yielding consistently lower errors in high-density regions.}
  \Description{Grid of 2D density field slices showing three test samples across columns: input observation at z=0, ground truth at z=127, diffusion reconstruction, Cosmo3DFlow reconstruction, and error maps for both methods. Dashed boxes highlight regions where Cosmo3DFlow preserves sharper features than the baseline.}%
  \label{fig:qualitative_results}
\end{figure*}
\subsection{Experimental Settings}

\noindent \textbf{Dataset.}
We evaluate on the following datasets from the Quijote Latin Hypercube (LH) simulations\footnote{\url{https://quijote-simulations.readthedocs.io/en/latest/LH.html}}~\cite{villaescusa-navarroQuijoteSimulations2020}, summarized in Table~\ref{tab:quijote_datasets}. All suites use a periodic box volume of ${(1000\,h^{-1}\,\text{Mpc})}^{3}$ with $512^3$ particles. For the Standard LH and Big Sobol Sequence (BSQ) sets, the five core $\Lambda$CDM parameters are varied. In the $f_{\text{NL}}^{\text{local}}$ set, these parameters are held fixed at the fiducial cosmology ($\Omega_m=0.3175, \Omega_b=0.049, h=0.6711, n_s=0.9624, \sigma_8=0.834$) to isolate the impact of primordial non-Gaussianity on large-scale structure. We collect the snapshots from the publicly available source at redshifts $z=127$ (initial conditions) and $z=0$ (present-day universe).

\begin{table}[!htbp]
  \centering
  \caption{Configurations of the Quijote simulation subsets used to train and evaluate our models.}%
  \label{tab:quijote_datasets}
  \resizebox{\columnwidth}{!}{%
  \begin{tabular}{lcll}
    \toprule
    \textbf{Dataset} & Samples & Parameters & Description \\
    \midrule
    SLH & 2,000 & 5 core $\Lambda$CDM & Baseline cosmology \\
    BSQ & 1,000 & 5 core $\Lambda$CDM & Extended sampling \\
    NG-LH & 1,000 & $f_{\text{NL}}^{\text{local}} \in [-300, 300]$ & Non-Gaussianity \\
    \bottomrule
  \end{tabular}%
  }
\end{table}

\noindent \textbf{Standard LH (SLH).} This suite comprises 2,000 full $N$-body dark matter simulations with cosmological parameters sampled uniformly via Latin-Hypercube sampling across $\Omega_m \in [0.1, 0.5]$, $\Omega_b \in [0.03, 0.07]$, $\sigma_8 \in [0.6, 1.0]$, $h \in [0.5, 0.9]$, and $n_s \in [0.8, 1.2]$. We partition the simulations into training, validation, and test sets of sizes 1,800, 100, and 100, respectively.

\noindent \textbf{Big Sobol Sequence (BSQ).} A large collection of $N$-body simulations designed for machine learning, varying the same 5 core $\Lambda$CDM parameters as the SLH but generated using a Sobol Sequence. We use the first 1,000 samples, split 8:1:1 for training, validation, and test.

\noindent \textbf{Non-Gaussian Local LH (NG-LH).} 1,000 simulations exploring the impact of primordial non-Gaussianities on large-scale structure by varying $f_{\text{NL}}^{\text{local}} \in [-300, 300]$ via Latin-Hypercube sampling, split 8:1:1.

\noindent \textbf{Data generation budget.} Because we utilize a subset of the publicly available Quijote suite~\cite{villaescusa-navarroQuijoteSimulations2020}, no local compute budget was required to generate the training data. For context, generating the original simulation suite required approximately 35 million CPU core-hours.

\noindent \textbf{Density Field Construction.} The 3D matter density fields are constructed from the $N$-body simulation snapshots using the Pylians library~\cite{2018ascl.soft11008V} following~\cite{leginPosteriorSamplingInitial2024}. For the late-time snapshots at $z=0$, we employ the Piecewise Cubic Spline mass assignment scheme to minimize aliasing and ensure a high-fidelity representation of the non-linear cosmic web. For the initial conditions at $z=127$, we use the Cloud-in-Cell (CIC) scheme, which is sufficient for the low-contrast fluctuations at high redshift. Each density field is computed at three resolutions ($32^3$, $64^3$, and $128^3$) and converted into the dimensionless overdensity field, $\delta = \rho/\bar{\rho} - 1$, to serve as the input and target for our generative models. We apply a logarithmic transformation to the $z=0$ observation fields due to their high dynamic range. Both fields are standardized afterwards for model training.

\noindent \textbf{Evaluation Metrics.} Pointwise error alone is insufficient for scientific use: a model can attain low VRMSE while failing to recover the power spectrum $P(k)$, so we additionally evaluate cosmological summary statistics that the reconstructed field must respect. We consider several metrics for evaluation, each serving a different purpose. All are implemented using the Pylians library~\cite{2018ascl.soft11008V}.
\begin{enumerate}
  \item \textbf{Variance-normalized RMSE (VRMSE).} The root mean squared error (RMSE) and its normalized variants are commonly used metrics to quantify the pointwise accuracy of simulations~\cite{rozetLostLatentSpace2026, ohanaWellLargeScaleCollection2024}. Formally, for two spatial fields $\mathbf{a}$ and $\mathbf{b}$, the VRMSE is defined as follows, where $\langle \cdot \rangle$ denotes the spatial mean operator, and $\epsilon = 10^{-6}$ is a numerical stability term:
\begin{equation}
  \text{VRMSE}(\mathbf{a}, \mathbf{b}) = \sqrt{ \frac{\langle {(\mathbf{a}-\mathbf{b})}^{2} \rangle}{ \langle{(\mathbf{a}-\langle \mathbf{a} \rangle)}^{2} \rangle + \epsilon}}
\end{equation}

\item \textbf{Power Spectrum $R^2$-score.} The predicted dark matter density distribution should remain similar to the ground truth. Following~\cite{riverosConditionalDiffusionFlowModels2025}, we calculate the $R^2$-score between the target 3D density field and the predicted samples. This shows how the model performs on the power spectrum.

\item \textbf{Cross-Correlation, $C(k)$.} This metric measures the scale-dependent correlation between reconstructed and true density fields, providing a metric insensitive to overall amplitude normalization~\cite{jascheBayesianPhysicalReconstruction2013, modiFlowPMDistributedTensorFlow2021}. It is computed using the following, where $P_{11}(k)$ and $P_{22}(k)$ denote the power spectra of the sample and truth, respectively:
\begin{equation}
  C(k) = \frac{P_{12}(k)}{\sqrt{P_{11}(k) \cdot P_{22}(k)}}
\end{equation}
\item \textbf{Transfer Function, $T(k)$.} The transfer function $T(k)$ characterizes the scale-dependent fidelity of reconstruction by measuring the ratio of recovered to true power at each spatial scale. We define the transfer function as: $T(k) = \sqrt{P_{\text{sample}}(k)/ P_{\text{target}}(k)}$.

\end{enumerate}

\subsection{Main Results}

\noindent \textbf{Baseline.} We use the score-based diffusion model from Legin et al.~\cite{leginPosteriorSamplingInitial2024} as the baseline generative model, as it represents the state-of-the-art on our target problem. We refer to it as the ``diffusion'' baseline. The U-Net parameters are kept the same as in our model to obtain a fixed latent dimension ($8 \times 8 \times 8$) across resolutions; hence, the compression ratio matches across resolutions. This ensures a fair comparison between the two models.

\begin{table*}[!tb]
  \centering
  \caption{Performance on Quijote datasets. Cosmo3DFlow (Ours, shaded grey) consistently outperforms the diffusion baseline across resolutions. Bold indicates substantial improvements.}%
  \label{tab:quijote}
  \begin{tabular}{ll *{4}{>{\columncolor{gray!20}}cc}}
    \toprule
    \multirow{2}{*}{Dataset} & \multirow{2}{*}{$N^3$} & \multicolumn{2}{c}{VRMSE$\downarrow$} & \multicolumn{2}{c}{Cross Correlation$\uparrow$} & \multicolumn{2}{c}{Power Spectrum $R^2$$\uparrow$} & \multicolumn{2}{c}{Transfer Function$\uparrow$}\\
    \cmidrule(lr){3-4} \cmidrule(lr){5-6} \cmidrule(lr){7-8} \cmidrule(lr){9-10}
    & & \textbf{Ours} & Diffusion & \textbf{Ours} & Diffusion  & \textbf{Ours} & Diffusion  & \textbf{Ours} & Diffusion \\ \midrule

    \multirow{3}{*}{\shortstack[l]{Standard\\Latin\\Hypercube}}
    & $128^3$ & \textbf{0.50} & 0.63 & 0.88 & 0.82 & \textbf{0.99} & 0.70 & \textbf{0.99} & 0.80  \\
    & $64^3$  & 0.47 & 0.68 & 0.92 & 0.89 & 0.98 & 0.59 & 0.98 & 0.59  \\
    & $32^3$  &  \textbf{0.25} & 0.82 &  0.98 & 0.85 &  0.97 & 0.48 &  0.96 & 0.48  \\
    \midrule

    \multirow{3}{*}{\shortstack[l]{Big Sobol\\Sequence}}
    & $128^3$ &  0.62 & 0.64 &  0.80 & 0.79 &  0.99 & 0.84 &  0.95 & 0.88  \\
    & $64^3$  &  0.53 & 0.65 &  0.88 & 0.88 &  0.98 & 0.83 &  0.94 & 0.81  \\
    & $32^3$  &  0.37 & 0.79 &  0.95 & 0.85 &  \textbf{0.95} & 0.48 &  0.94 & 0.71  \\
    \midrule

    \multirow{3}{*}{\shortstack[l]{Non-Gauss\\($f_\text{NL}$ LH)}}
    & $128^3$ &  0.56 & 0.59 &  0.86 & 0.83 &  1.00 & 1.00 &  0.98 & 0.98 \\
    & $64^3$  &  0.47 & 0.57 &  0.93 & 0.89 &  1.00 & 1.00 &  0.99 & 0.99 \\
    & $32^3$  &  0.31 & 0.67 &  0.97 & 0.87 &  1.00 & 0.98 &  0.99 & 0.98 \\
    \bottomrule
  \end{tabular}
\end{table*}

\begin{figure}[!htbp]
  \centering
  \begin{subfigure}{\linewidth}
    \includegraphics[width=\linewidth]{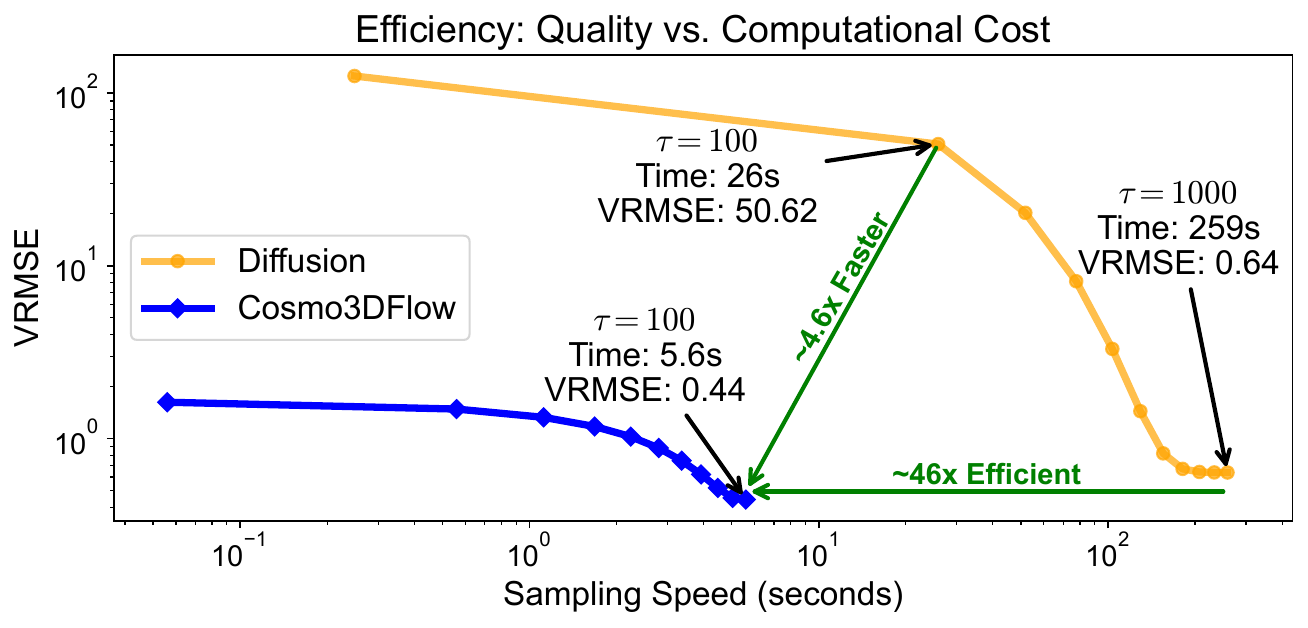}
    \caption{Efficiency vs.\ quality.}
    \Description{Scatter plot comparing computational efficiency (sampling time) versus reconstruction quality (VRMSE) for the baseline and Cosmo3DFlow models at different numbers of sampling steps.}%
    \label{fig:efficiency}
  \end{subfigure}

  \vspace{0.8em}
  \begin{subfigure}{\linewidth}
    \includegraphics[width=\linewidth]{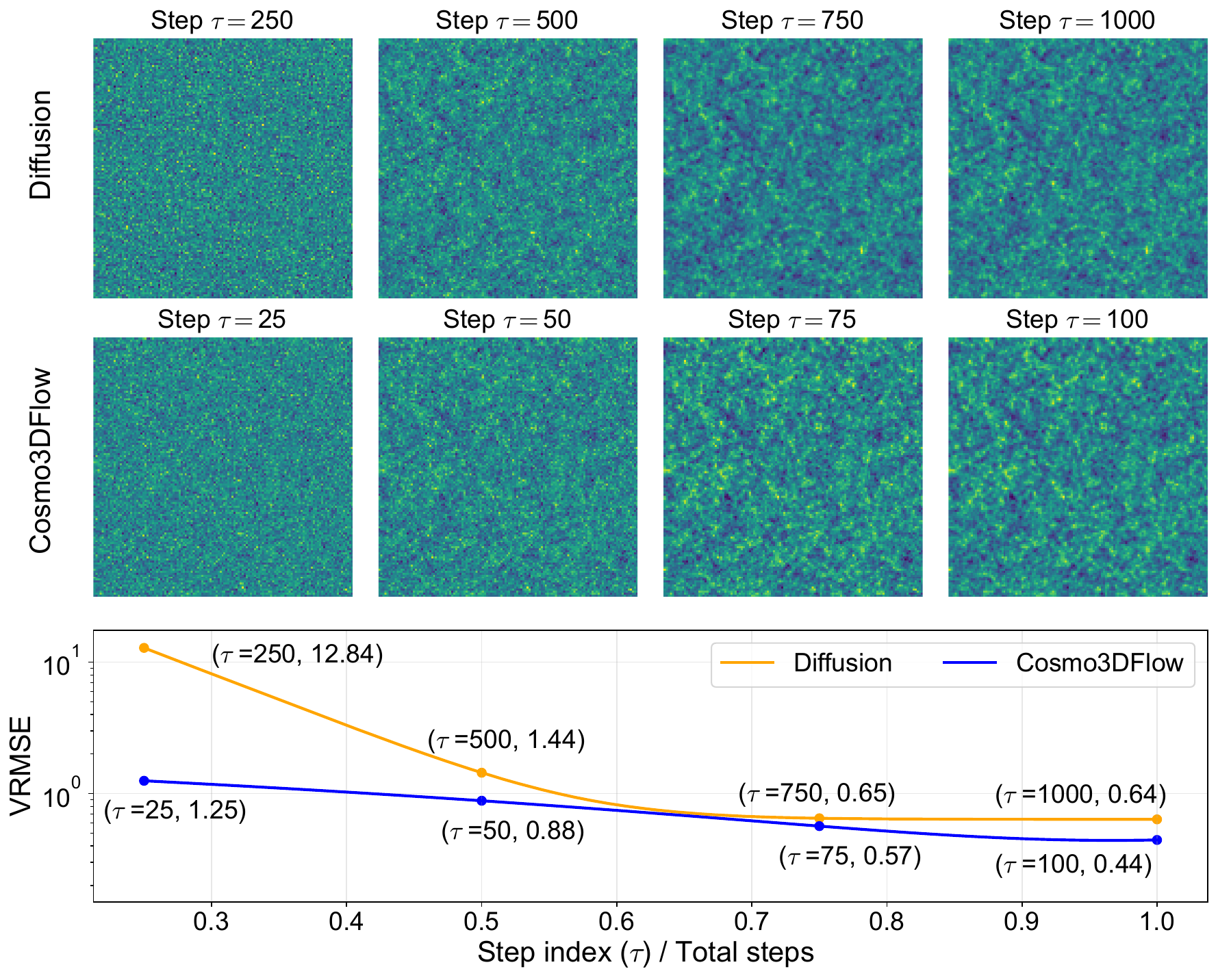}
    \caption{Sample generation from Gaussian noise.}
    \Description{Grid showing the progressive denoising and generation process from Gaussian noise to reconstructed density fields, comparing diffusion and our model across multiple time steps, with VRMSE loss values shown in the bottom row.}%
    \label{fig:convergence_steps}
  \end{subfigure}

  \vspace{0.8em}
  \begin{subfigure}{\linewidth}
    \includegraphics[width=\linewidth]{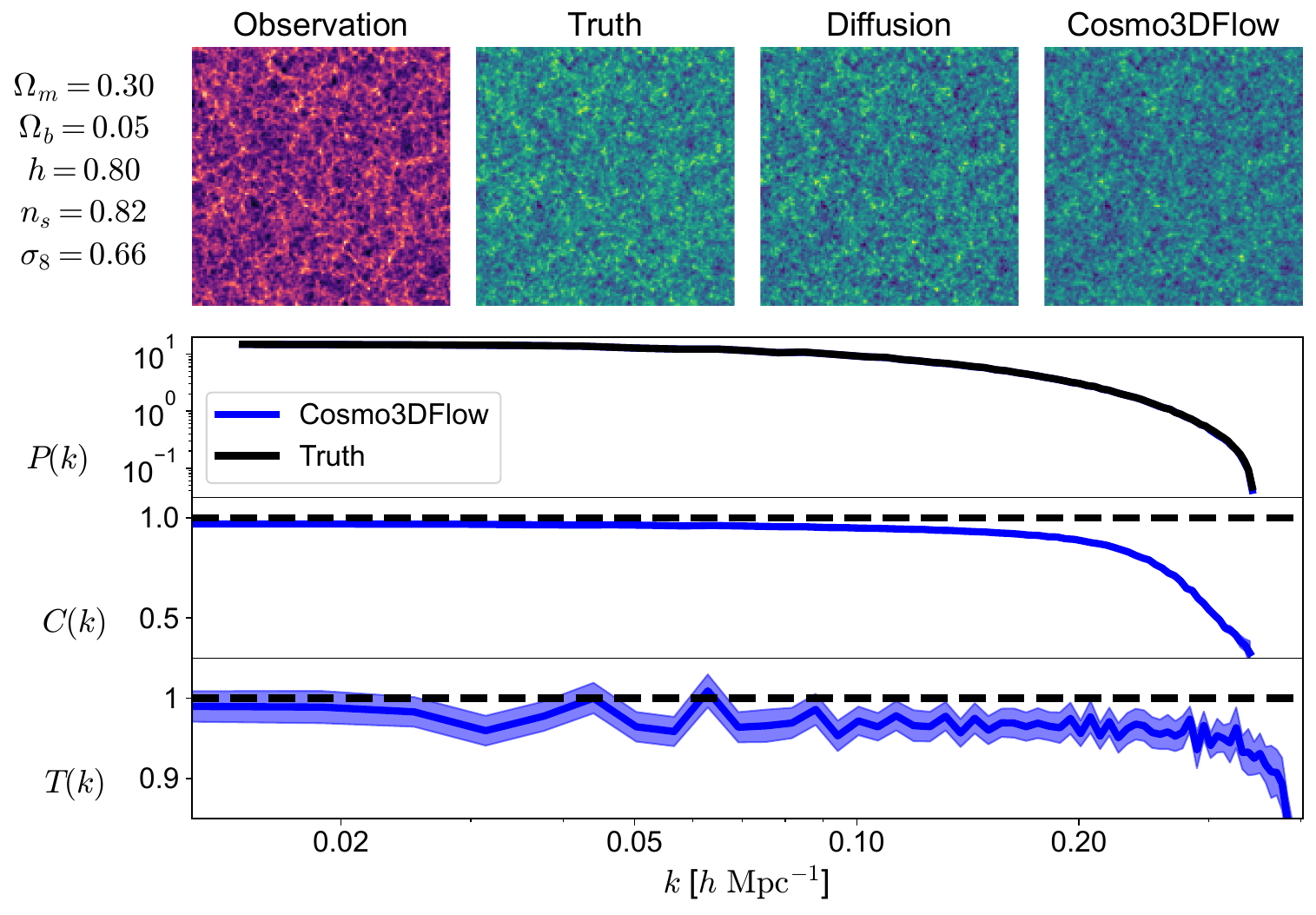}
    \caption{Quantitative evaluation on the Standard LH, resolution $128^3$.}
    \Description{Multi-panel figure showing density field reconstructions in the top row and three physics validation metrics below: power spectrum P(k), cross-correlation coefficient C(k), and transfer function, all plotted against wavenumber k on a log scale.}%
    \label{fig:qualitative_metrics}
  \end{subfigure}

  \caption{\textbf{Quantitative Results.} \textbf{(a)} Cosmo3DFlow achieves $4.6\times$ lower per-step cost. \textbf{(b)} VRMSE progression shows significantly faster convergence. \textbf{(c)} Near-perfect alignment with the truth across power spectrum, cross-correlation, and transfer function. Note the logarithmic scale on the horizontal axis.}
\end{figure}

\noindent \textbf{Comparison to the diffusion baseline.} We assess our method against the diffusion baseline along four axes. Cosmo3DFlow preserves sharp small-scale structure without the blurring seen in the baseline (Figure~\ref{fig:qualitative_results}). Quantitatively, it reaches matching reconstruction quality with substantially fewer sampling steps (Figure~\ref{fig:efficiency}), converges faster in VRMSE (Figure~\ref{fig:convergence_steps}), and aligns near-perfectly with the truth on power spectrum, cross-correlation, and transfer function (Figure~\ref{fig:qualitative_metrics}). As summarized in Table~\ref{tab:quijote}, Cosmo3DFlow consistently outperforms the diffusion baseline across all datasets and resolutions. At $128^3$ on the Standard Latin Hypercube, it achieves a VRMSE of 0.50 vs.\ 0.63, cross-correlation of 0.88 vs.\ 0.82, and a power spectrum $R^2$-score of 0.99 vs.\ 0.70. These gains are consistent across the Big Sobol Sequence and Non-Gaussian datasets, demonstrating robustness to varying cosmological parameters and primordial non-Gaussianity. This baseline uses a stochastic differential equation (SDE) for diffusion sampling. To ensure an equitable comparison, we further isolate the contribution of deterministic ODE sampling from our wavelet-space formulation and additionally compare Cosmo3DFlow (ODE) against the deterministic Probability Flow ODE (PF-ODE) variant of the diffusion baseline using the Euler method, as shown in Table~\ref{tab:pfode} of the Appendix. As reported on the Quijote Standard Latin Hypercube, Cosmo3DFlow still maintains a sizable margin in speed and quality.

Appendix~\ref{sec:appendix_posterior} reports posterior calibration and diversity for our model.

\subsection{Generalization to Realistic Observational Tracers}
To demonstrate that the Cosmo3DFlow model works not only on dark-matter density fields, but also on realistic observational inputs, we also train and evaluate on the Quijote Standard \textit{halo dataset} at $128^3$, where halos act as sparse, biased, and noisy tracers that mimic actual telescope survey data (e.g., DESI or Euclid surveys). Because the true initial conditions of the real universe are unobservable, halos---the gravitationally bound structures that host galaxies and galaxy clusters---provide the most faithful ground-truth signal available. Prior work~\cite{huangCosmoBenchMultiscaleMultiview2025,parkerInitialConditionsGalaxies2025a} follows the same convention of training and evaluating on simulated datasets. As shown in Table~\ref{tab:halo}, Cosmo3DFlow remains robust under this noisier and sparser conditioning, outperforming the diffusion baseline on all four metrics.

\begin{table}[!htbp]
  \centering
  \caption{Evaluation on the Quijote Standard halo dataset ($128^3$), where halos act as sparse, biased tracers of the underlying density. \textbf{Bold} indicates the best in each column.}%
  \label{tab:halo}
  \begin{tabular}{lcccc}
    \toprule
    Model & VRMSE$\downarrow$ & \shortstack{Cross\\Correlation}$\uparrow$ & $R^2$$\uparrow$ & \shortstack{Transfer\\Function}$\uparrow$ \\ \midrule
    Diffusion & 1.08 & 0.55 & 0.85 & 0.85 \\
   \textbf{Ours} & \textbf{0.93} & \textbf{0.56} & \textbf{0.90} & \textbf{0.91} \\
    \bottomrule
  \end{tabular}
\end{table}

\subsection{Ablation Study}
\noindent \textbf{Component ablation.} We evaluate the impact of our added components on model performance at $32^3$ resolution using the Standard Latin Hypercube set. For a fair comparison, Voxel FM (without wavelets) utilizes an additional downsampling/upsampling layer to maintain the same compression ratio with a latent dimension $8 \times 8 \times 8$ as the wavelet-based configurations. Table~\ref{tab:ablation} shows that Voxel FM already achieves good performance. Naively applying wavelet flow matching, without our scale-aware architectural interventions, yields a worse VRMSE than Voxel FM (0.34 vs.\ 0.26), indicating that simply stacking wavelets onto flow matching is insufficient on highly sparse cosmological data. Adding Scale Conditioning and Cross-Scale Skips bridges the gap between the spatial and wavelet domains, providing the necessary context to resolve features across different scales, and the power-spectrum weighting balances reconstruction error across wavenumbers. We omitted the cross-correlation because all $32^3$ models yielded approximately 0.98.

\begin{table}[!htbp]
  \centering
  \caption{Ablation study of model components evaluated on the Standard Latin Hypercube, showing that the added components have a positive impact on the overall performance. \textbf{Bold} indicates the best in each column.}
  \label{tab:ablation}
  \begin{tabular}{ccccccc}
    \toprule
    \multicolumn{4}{c}{\textbf{Components}} & \multicolumn{3}{c}{\textbf{Evaluation Metrics}} \\ \cmidrule(lr){1-4} \cmidrule(lr){5-7}
    Wave & Cond & Skip & $P(k)$ & VRMSE$\downarrow$ & $R^2$$\uparrow$ & $T(k)$$\uparrow$ \\ \midrule
    $\times$    & $\times$    & $\times$    & $\times$    & 0.26          & 0.95          & 0.94          \\
    \checkmark  & $\times$    & $\times$    & $\times$    & 0.34          & 0.95          & 0.95          \\
    \checkmark  & \checkmark  & \checkmark  & $\times$    & 0.27          & 0.96          & 0.96 \\
    \grayrow \checkmark  & \checkmark  & \checkmark  & 0.01        & \textbf{0.25} & \textbf{0.97} & \textbf{0.96} \\
    \bottomrule
  \end{tabular}\\[2pt]
  \parbox{\columnwidth}{\small\textit{Components:} Wave\,=\,wavelet decomposition; Cond\,=\,scale conditioning; Skip\,=\,cross-scale skips; $P(k)$\,=\,power-spectrum weighting. \textit{Metrics:} VRMSE, power-spectrum $R^2$, and transfer function $T(k)$.}
\end{table}

\subsection{Complexity and Scaling Analysis}

\begin{table}[!t]
  \centering
  \caption{Wavelet-space contribution: Sampling efficiency of Cosmo3DFlow vs.\ standard voxel-space flow matching (Voxel FM) and diffusion.}%
  \label{tab:performance_comparison}
  \begin{tabular}{llrr>{\columncolor{gray!20}}rr}
    \toprule
    $N^3$ & Metric & Diffusion & Voxel FM & \textbf{Ours} & Gain \\
    \midrule
    $32^3$  & Time (s)   & 10.8 & 1.32 & 1.06 & $10.2\times$  \\
            & Memory (GB) & 0.22  & 0.37  & 0.21  & $4.5\%$      \\
    \midrule
    $64^3$  & Time (s)   & 39.2  & 3.51 & 1.52 & $25.8\times$ \\
            & Memory (GB) & 0.80  & 0.81  & 0.37  & $53.8\%$      \\
    \midrule
    $128^3$ & Time (s)   & 259  & 23.7 & 5.60 & $46.2\times$  \\
            & Memory (GB) & 3.93 & 5.47 & 1.74 & $55.7\%$      \\
    \bottomrule
  \end{tabular}\\[2pt]
  \parbox{\columnwidth}{\small\textit{Gain:} relative improvement against Diffusion (Speedup for Time, \% Less for Memory).}
\end{table}

\begin{table}[!t]
  \centering
  \caption{Per-component theoretical complexity of Cosmo3DFlow (Ours) vs.\ the diffusion baseline.}%
  \label{tab:complexity}
  \begin{tabular}{lcc}
    \toprule
    Component & Diffusion & Ours \\
    \midrule
    Asymptotic scaling  & $\mathcal{O}(N^3)$     & $\mathcal{O}(N^3)$ \\\midrule
    Spatial domain      & $N^3$                  & $N^3/8$ \\\midrule
    Integration steps   & $K'{=}1{,}000$         & $K{=}100$ \\\midrule
    Sampling complexity & $\mathcal{O}(K' N^3 C)$ & $\mathcal{O}(K N^3 C / 8)$ \\
    \bottomrule
  \end{tabular}\\[2pt]
  \parbox{\columnwidth}{\small\textit{Symbols:} $N$\,=\,grid size; $C$\,=\,channels; $K, K'$\,=\,sampling step counts.}
\end{table}

\begin{table}[!t]
  \centering
  \caption{Cosmo3DFlow at $256^3$ on Standard Latin Hypercube.}
  \label{tab:scaling256}
  \begin{tabular}{lcccc}
    \toprule
    Setup & VRMSE$\downarrow$ & $C(k)$$\uparrow$ & $R^2$$\uparrow$ & $T(k)$$\uparrow$ \\
    \midrule
    $128^3$        & 0.50          & 0.88          & 0.99 & 0.99 \\
    \midrule
    $256^3$        & 0.85          & 0.55          & 0.97 & 0.95 \\
    \grayrow $256^3$+TL$^\dagger$   & 0.68 & 0.72 & 0.96 & 0.92 \\
    \bottomrule
  \end{tabular}\\[2pt]
  \parbox{\columnwidth}{\small\textit{Metrics:} VRMSE, cross-correlation $C(k)$, power spectrum $R^2$, and transfer function $T(k)$. $^\dagger$ denotes transfer learning (TL) from the $128^3$ checkpoint.}
\end{table}

\noindent \textbf{Memory and runtime efficiency.} The Cosmo3DFlow framework performs computations within a decomposed coefficient space rather than the high-resolution spatial domain. As shown in Table~\ref{tab:performance_comparison}, this mitigates the cubic scaling bottlenecks inherent in $128^3$ volumes, achieving a 68.2\% reduction in peak memory compared to standard flow matching and a 55.7\% reduction relative to the diffusion baseline at $128^3$. The wavelet decomposition thus preserves structural fidelity while lowering computational overhead; the per-factor speed breakdown is detailed below.

\noindent \textbf{Theoretical ceiling vs.\ practical efficiency.} Both approaches share the same $\mathcal{O}(N^3)$ asymptotic upper bound; our model's acceleration comes from reducing the constant factors. Composing the per-step $8\times$ spatial-domain factor in Table~\ref{tab:complexity} with the $10\times$ reduction in integration steps yields an ${\approx}80\times$ theoretical end-to-end efficiency gain, against the $46\times$ empirical wall-clock observed at $128^3$ (Table~\ref{tab:performance_comparison}). The remaining gap is dominated by 3D convolutions being memory-bandwidth-bound rather than FLOPs-bound on the target hardware, together with fixed costs that do not shrink with $N$ (DWT/IDWT, time embedding, kernel launches); accordingly the observed acceleration grows from $10.2\times$ at $32^3$ to $25.8\times$ at $64^3$ to $46.2\times$ at $128^3$, converging toward the theoretical ceiling as $N$ increases.

\noindent \textbf{Scaling to $\mathbf{256^3}$.} Our model extends naturally to the $256^3$ full volume, a regime in which the diffusion baseline runs out of memory (projected ${\sim}150$\,GB VRAM for training, extrapolated from the $128^3$ row of Table~\ref{tab:performance_comparison}). Training Cosmo3DFlow at $256^3$ requires 23\,GB of VRAM (batch size 2) and 40 GPU-hours on a single A100 for 100 epochs over 1,800 samples. At inference, each posterior sample takes about 41 seconds and 13\,GB of VRAM\@. Transfer learning using the $128^3$ checkpoint substantially accelerates convergence and improves accuracy, as shown in Table~\ref{tab:scaling256}. The diffusion baseline cannot be trained at $256^3$ due to out-of-memory errors.

\section{Conclusion}

We introduced Cosmo3DFlow, a generative framework combining flow matching with wavelet-space modeling for high-dimensional inverse problems. When density fields are transformed into the wavelet domain, voids collapse into near-zero high-frequency coefficients while informative structures are preserved compactly---a physically informed exploitation of \textit{wavelet sparsity}. This achieves up to \textbf{$46\times$ faster} sampling than diffusion baselines at $128^3$ resolution, as wavelet decomposition naturally aligns with multi-scale cosmological structure, enabling sparse velocity fields that require far fewer ODE integration steps.

As cosmological simulations scale to larger volumes and finer resolutions ($512^3$, $1024^3$), the voxel void fraction will remain constant or increase (as we resolve smaller voids), but the computational cost of processing it on a grid will scale cubically ($N^3$). Wavelet sparsity mitigates this scaling bottleneck. It ensures that computational resources are allocated to the structure of the universe---the mass, the energy, and the information---rather than its emptiness. This shift from spatial to spectral modeling represents not just an optimization, but a necessary evolution for the next generation of high-dimensional astrophysical inference.


This work provides valuable insights into mapping sparse spatial structures to the frequency domain, and the proposed Wavelet Flow Matching paradigm enables more stable, efficient ODE solvers with larger steps while preserving fine detail. We believe the core algorithmic innovation of this paper---translating spatial emptiness into spectral sparsity---can be generalized to other fields such as medical imaging (MRI/CT), by enabling high-resolution synthesis; computational fluid dynamics, by focusing resources on turbulent regions; and meteorology, by improving global forecasting resolution while bypassing grid-scale computational limits.


\section*{Limitations and Ethical Considerations}

While our method proves robust on both dark matter density fields and halo catalogs derived from $N$-body simulations, full hydrodynamical simulation suites (incorporating baryonic physics) and alternative wavelet families remain to be explored. The synthetic simulation data poses no privacy or ethical concerns.

\section*{GenAI Disclosure}

Large Language Models (LLMs) were used for the initial drafting of figures and text polishing.

\begin{acks}
  We acknowledge support from the \grantsponsor{NSF}{National Science Foundation}{https://www.nsf.gov/} under Cooperative Agreement \grantnum{NSF}{2421782}, the \grantsponsor{SIMONS}{Simons Foundation}{https://www.simonsfoundation.org/} through award \grantnum{SIMONS}{MPS-AI-00010515} and NSF Simons Cosmic AI Seed Grant \grantnum{SIMONS}{AWD-006703}. We are grateful for the resources and support provided by UVA Research Computing.
\end{acks}

\bibliographystyle{ACM-Reference-Format}
\balance
\bibliography{bibliography}

\appendix

\section{3D U-Net Architecture Details}%
\label{sec:appendix_unet}

\begin{figure}[!htbp]
  \centering
  \includegraphics[width=\columnwidth]{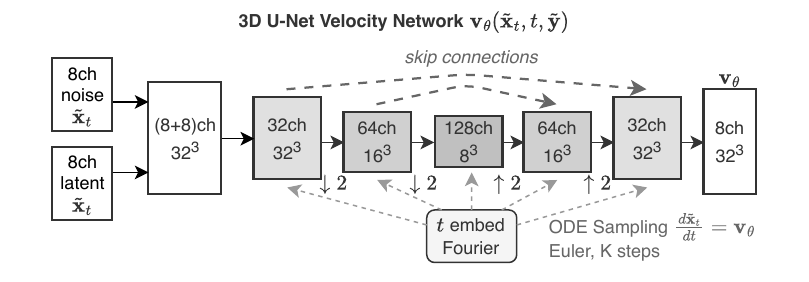}
  \caption{3D U-Net velocity network at $64^3$. It predicts wavelet-space velocity from noisy coefficients and conditioning. The architecture features time-embeddings, skip connections, and an $8^3$ latent bottleneck. See Algorithm~\ref{alg:unet} for details.}
  \label{fig:3dunet}
  \Description{Detailed architecture diagram of the 3D U-Net velocity network showing the encoder-decoder structure with skip connections, residual blocks, time embedding injection, and the flow of volumetric data through the network.}
\end{figure}

Figure~\ref{fig:3dunet} illustrates the detailed architecture of our 3D U-Net velocity network. The network follows an encoder-decoder structure specifically designed for volumetric cosmological data processing in wavelet space. We provide more details in Algorithm~\ref{alg:unet}.

\noindent \textbf{Input Processing.} The network receives a 16-channel input by concatenating noisy wavelet coefficients
$\tilde{\mathbf{x}}_t \in \mathbb{R}^{8 \times (N/2)^3}$ with the conditioning observation $\tilde{\mathbf{y}} \in \mathbb{R}^{8 \times (N/2)^3}$. An initial $3 \times 3 \times 3$ convolution projects this to the base feature dimension $n_f = 32$.

\noindent \textbf{Encoder Structure.} The encoder consists of 2, 3, and 4 downsampling layers for the image resolutions 32, 64, and 128, respectively. Each level contains $N_{\text{res}} = 2$ residual blocks that incorporate time conditioning through additive feature modulation. Spatial downsampling between levels is performed via strided average pooling with a factor of 2.

\noindent \textbf{Residual Block.} The residual blocks employ the BigGAN-style architecture:
\begin{equation}
  \mathbf{h}' = \text{Conv}(\text{SiLU}(\text{GN}(\mathbf{h}))) + \text{Dense}(\text{SiLU}(\mathbf{e}_t))
\end{equation}
\begin{equation}
  \mathbf{h}'' = \text{Conv}(\text{Dropout}(\text{SiLU}(\text{GN}(\mathbf{h}'))))
\end{equation}
\begin{equation}
  \text{ResBlock}(\mathbf{h}, \mathbf{e}_t) = (\mathbf{h}'' + \text{Skip}(\mathbf{h})) / \sqrt{2}
\end{equation}
where GN denotes Group Normalization with $\min(c/4, 32)$ groups ($c$ being the number of channels), and Skip denotes a $1 \times 1 \times 1$ convolution when channel dimensions change.

\noindent \textbf{Bottleneck.} At the coarsest resolution, the network processes features at the fixed $8^3$ bottleneck resolution with 256 channels, capturing global context essential for coherent large-scale structure generation.

\noindent \textbf{Decoder Structure.} The decoder symmetrically mirrors the encoder, with each level receiving skip connections from the corresponding encoder stage. Features are upsampled via nearest-neighbor interpolation followed by convolution, then concatenated with skip features before processing through residual blocks. This design preserves fine-grained spatial information while incorporating global context.

\noindent \textbf{Time Conditioning.} The continuous time variable $t \in [0, 1]$ is embedded using Gaussian Fourier features:
\begin{equation}
  \gamma(t) = \left[\sin(2\pi \mathbf{W} t), \cos(2\pi \mathbf{W} t)\right] \in \mathbb{R}^{2d}
\end{equation}
where $\mathbf{W} \in \mathbb{R}^d$ are learnable frequency parameters initialized from $\mathcal{N}(0, \sigma^2)$ with $\sigma = 16$. This embedding is projected through two dense layers with SiLU activations to produce $\mathbf{e}_t \in \mathbb{R}^{4 n_f} = \mathbb{R}^{128}$, which is injected into each residual block via a dense projection added to the intermediate features.

\noindent \textbf{Output Layer.} The final output is produced by Group Normalization, SiLU activation, and a $3 \times 3 \times 3$ convolution projecting to 8 output channels (predicted wavelet-space velocity).

\begin{algorithm}[!htb]
  \caption{\textit{U-Net Forward Pass with Scale Conditioning and Cross-Scale Skips}}\label{alg:unet}
  \setlength{\fboxsep}{1pt}%
  \begin{algorithmic}[1]
    \STATE \textbf{Input:} Noisy wavelet coefficients $\tilde{\mathbf{x}}_t$, flow time $t$, conditioning observation $\tilde{\mathbf{y}}$
    \STATE \textbf{Output:} Predicted wavelet-space velocity $\mathbf{v}_t$

    \STATE Stack $[\tilde{\mathbf{x}}_t;\tilde{\mathbf{y}}]$ as input; encode $t$ to embedding $\mathbf{e}_t$; project to feature space

    \STATE \textit{\textbf{Stage 1: Encoder with scale conditioning}}
    \FOR{each encoder level $i = 0$ to $\mathcal{L}{-}1$}
      \STATE Apply residual blocks guided by $\mathbf{e}_t$; cache as U-Net skip
      \STATE {\parbox{0.95\linewidth}{Scale conditioning: inject $\tilde{\mathbf{y}}$ at level-$i$ resolution}}
      \STATE {\parbox{0.95\linewidth}{Cross-scale skip: snapshot encoder features $\mathbf{s}_i$}}
      \STATE Downsample to next coarser resolution (if $i < \mathcal{L}{-}1$)
    \ENDFOR

    \STATE \textit{\textbf{Stage 2: Bottleneck}}
    \STATE Integrate global structure at the coarsest resolution

    \STATE \textit{\textbf{Stage 3: Decoder with cross-scale fusion}}
    \FOR{each decoder level $i = \mathcal{L}{-}1$ down to $0$}
      \STATE Fuse U-Net skip; apply residual blocks guided by $\mathbf{e}_t$
      \STATE {\parbox{0.95\linewidth}{Cross-scale skip: add encoder snapshot $\mathbf{s}_i$}}
      \STATE Upsample to next finer resolution (if $i > 0$)
    \ENDFOR

    \STATE Project to 8-channel wavelet velocity $\mathbf{v}_t$ via GroupNorm, SiLU, and $3{\times}3{\times}3$ convolution
    \STATE \textbf{Return:} $\mathbf{v}_t$
  \end{algorithmic}
\end{algorithm}

\begin{table}[!b]
  \centering
  \caption{Posterior metrics across resolutions and noise scales. Coverage (PICP) remains high, while variance scales appropriately with noise level and finer structures (high $k$).}%
  \label{tab:posterior}
  \begin{tabular}{lcccccc}
    \toprule
    \multirow{2}{*}{$N^3$} & \multirow{2}{*}{Noise} & \multirow{2}{*}{PICP} & \multirow{2}{*}{\shortstack[c]{Relative\\Sample\\Variance}} & \multicolumn{3}{c}{Variance} \\
    \cmidrule(lr){5-7}
    & & & & low $k$ & mid $k$ & high $k$ \\ \addlinespace[2pt] \midrule
    \multirow{3}{*}{$32^3$}
    & 0.1 & 0.97 & 0.12 & 0.04 & 0.05 & 0.10 \\
    & 0.2 & 0.99 & 0.32 & 0.08 & 0.09 & 0.21 \\
    & 0.3 & 0.98 & 0.56 & 0.14 & 0.16 & 0.35 \\ \midrule
    \multirow{3}{*}{$64^3$}
    & 0.1 & 0.93 & 0.13 & 0.05 & 0.06 & 0.16 \\
    & 0.2 & 0.95 & 0.23 & 0.07 & 0.09 & 0.26 \\
    & 0.3 & 0.95 & 0.38 & 0.12 & 0.14 & 0.40 \\ \midrule
    \multirow{3}{*}{$128^3$}
    & 0.1 & 0.87 & 0.15 & 0.07 & 0.08 & 0.30 \\
    & 0.2 & 0.83 & 0.17 & 0.08 & 0.09 & 0.32 \\
    & 0.3 & 0.87 & 0.23 & 0.10 & 0.10 & 0.40 \\
    \bottomrule
  \end{tabular}
\end{table}

\section{Posterior Diversity, Calibration, and Uncertainty}%
\label{sec:appendix_posterior}

To address potential concerns about mode-seeking behavior in our conditional generator, we evaluate the Cosmo3DFlow posterior using three complementary metrics across three different inference noise scales (by default, $0.1\times$ Gaussian noise is added to the input condition for all models):
\begin{itemize}
  \item \emph{Prediction Interval Coverage Probability (PICP):} the fraction of spatial locations at which the ground truth falls within the predicted credible interval.
  \item \emph{Relative sample variance:} the sample variance of the generated posterior normalized by the variance of the true field.
  \item \emph{Diversity power spectrum:} a scale-dependent Fourier-space measure of how sample variance, normalized by the true power, behaves as a function of wavenumber $k$. We partition the power spectrum into three equal bands in $\log k$ space (large, medium, and small structures).
\end{itemize}

Results are reported in Table~\ref{tab:posterior}. Across resolutions and noise scales, Cosmo3DFlow exhibits excellent coverage (83--99\%) and healthy sample diversity (12--56\%). The variance is consistently $2.5\text{--}4.3\times$ larger at small scales (high $k$) than at large scales, which is physically expected: small-scale primordial features are largely erased by non-linear gravitational collapse and are therefore intrinsically harder to constrain from the $z=0$ observation, so the posterior correctly expresses larger uncertainty.

\section{Sampling Speed: SDE vs.\ PF-ODE vs.\ Cosmo3DFlow}

To isolate the contribution of deterministic ODE sampling from our wavelet-space formulation, we additionally compare Cosmo3DFlow against the deterministic Probability Flow ODE (PF-ODE) variant of the diffusion baseline using the Euler method. To ensure a fair comparison, both our ODE and the PF-ODE use 100 sampling steps; only the diffusion SDE retains its standard 1,000 steps. Results on the Quijote Standard Latin Hypercube are reported in Table~\ref{tab:pfode}. PF-ODE substantially accelerates diffusion sampling, but Cosmo3DFlow still maintains a sizable margin in speed and quality. At $128^3$, our model outperforms across all metrics and remains over $4\times$ faster than PF-ODE ($5.60$\,s vs.\ $23.9$\,s) while using less than half the GPU memory.

\begin{table*}[!t]
  \centering
  \caption{Sampler contribution: Sampling speed and quality on the Standard Latin Hypercube. \textbf{Bold} indicates the best in each column within each resolution block.}%
  \label{tab:pfode}
  \begin{tabular}{llrrrrrr}
    \toprule
    \multirow{2}{*}{$N^3$} & \multirow{2}{*}{Method} & \multicolumn{2}{c}{Efficiency} & \multicolumn{4}{c}{Quality} \\
    \cmidrule(lr){3-4} \cmidrule(lr){5-8}
    & & Memory (GB) & Time (s) & VRMSE$\downarrow$ & $C(k)$$\uparrow$ & $R^2$$\uparrow$ & $T(k)$$\uparrow$ \\ \midrule
    \multirow{3}{*}{$32^3$}
    & Diffusion (SDE)            & 0.22          & 10.8          & 0.82          & 0.85          & 0.48          & 0.48          \\
    & Diffusion (PF-ODE)         & 0.22          & 1.44          & 0.85          & 0.84          & 0.57          & 0.70          \\
    \grayrow & \textbf{Cosmo3DFlow (ODE)} & \textbf{0.21} & \textbf{1.06} & \textbf{0.25} & \textbf{0.98} & \textbf{0.97} & \textbf{0.96} \\ \midrule
    \multirow{3}{*}{$64^3$}
    & Diffusion (SDE)            & 0.80          & 39.2         & 0.68          & 0.89          & 0.59          & 0.59          \\
    & Diffusion (PF-ODE)         & 0.80          & 3.96          & 0.65          & 0.89          & 0.69          & 0.79          \\
    \grayrow & \textbf{Cosmo3DFlow (ODE)} & \textbf{0.37} & \textbf{1.52} & \textbf{0.47} & \textbf{0.92} & \textbf{0.98} & \textbf{0.98} \\ \midrule
    \multirow{3}{*}{$128^3$}
    & Diffusion (SDE)            & 3.93          & 259        & 0.63          & 0.82          & 0.70          & 0.80          \\
    & Diffusion (PF-ODE)         & 3.93          & 23.9         & 0.62          & 0.85          & 0.46          & 0.84          \\
    \grayrow & \textbf{Cosmo3DFlow (ODE)} & \textbf{1.74} & \textbf{5.60} & \textbf{0.50} & \textbf{0.88} & \textbf{0.99} & \textbf{0.99} \\
    \bottomrule
  \end{tabular}\\[2pt]
  \parbox{\textwidth}{\small\textit{Samplers:} SDE\,=\,standard stochastic diffusion sampler (1,000 steps); PF-ODE\,=\,deterministic probability-flow variant (100 steps); Cosmo3DFlow\,=\,100 steps.}
\end{table*}

\section{Notations}

Table~\ref{tab:notation} summarizes the notation and key definitions used throughout this paper.

\begin{table}[!htbp]
  \centering
  \caption{Notation and key definitions used.}%
  \label{tab:notation}
  \resizebox{\columnwidth}{!}{%
  \begin{tabular}{llll}
    \toprule
    Symbol & Definition & Symbol & Definition \\
    \midrule
    \multicolumn{4}{l}{\textit{Problem Setup}} \\
    $\mathbf{x}$ & Initial conditions ($z{=}127$) & $\mathbf{y}$ & Observations ($z{=}0$) \\
    $\hat{\mathbf{x}}$ & Recovered initial-condition field & $z$ & Cosmological redshift \\
    $N$ & Grid resolution per dim & $p(\mathbf{x}|\mathbf{y})$ & Posterior of initial conds \\
    $\delta$ & Density contrast $\rho/\bar{\rho}-1$ & $\rho,\bar{\rho}$ & Density and mean density \\
    \midrule
    \multicolumn{4}{l}{\textit{Flow Matching}} \\
    $t$ & Flow time $\in [0,1]$ & $\mathbf{x}_0$ & Prior (Gaussian) sample \\
    $\mathbf{x}_1$ & Target data sample & $\mathbf{x}_t$ & $t\mathbf{x}_1 + (1{-}t)\mathbf{x}_0$ \\
    $\mathbf{u}_t$ & Target velocity $\mathbf{x}_1{-}\mathbf{x}_0$ & $\mathbf{v}_\theta$ & Learned velocity network \\
    $K,K'$ & ODE integration step counts & $\theta$ & Network parameters \\
    $\Delta t$ & ODE step size & $\mathrm{Lip}(\mathbf{v}_\theta)$ & Lipschitz constant of $\mathbf{v}_\theta$ \\
    \midrule
    \multicolumn{4}{l}{\textit{Wavelet Transform}} \\
    DWT & Discrete Wavelet Transform & IDWT & Inverse DWT \\
    $\tilde{\mathbf{x}},\tilde{\mathbf{y}}$ & Wavelet-transformed fields & $\mathbf{c}_{LLL}$ & Approximation coefficients \\
    $\mathbf{c}_{LLH},\ldots,\mathbf{c}_{HHH}$ & 7 detail bands & $L,H$ & Low/high-pass 1D filters \\
    \midrule
    \multicolumn{4}{l}{\textit{Training Objectives}} \\
    $\mathcal{L}_{\text{flow}}$ & Voxel-space FM loss & $\tilde{\mathcal{L}}_{\text{flow}}$ & Wavelet-space FM loss \\
    $\mathcal{L}_{\text{PS}}$ & Power spectrum loss & $\lambda_{\text{PS}}$ & Power spectrum weight \\
    $\mathcal{L}_{\text{total}}$ & Combined training objective & & \\
    \midrule
    \multicolumn{4}{l}{\textit{Evaluation Metrics}} \\
    VRMSE & Variance-normalized RMSE & $k,\mathbf{k}$ & Wavenumber (scalar/vector) \\
    $P(k)$ & Isotropic power spectrum & $C(k)$ & Cross-correlation at $k$ \\
    $T(k)$ & Transfer function & PICP & Pred.\ Interval Coverage Prob. \\
    $\epsilon$ & Numerical stabilizer & $\langle\cdot\rangle$ & Spatial mean operator \\
    \bottomrule
  \end{tabular}}
\end{table}

\end{document}